\documentclass[epj]{svjour}
\usepackage{graphics}
\usepackage{amssymb}
\usepackage{amsmath}
\def\scr{\rm\scriptscriptstyle }

\begin{document}

\title{Improved WKB approximation for quantum tunneling: Application to heavy ion fusion}
\author{A. J. Toubiana\inst{1,2,3}, L. F. Canto\inst{4,5}, M. S. Hussein\inst{6,7,8}}
\institute{$^1$ Departamento de Engenharia Nuclear, Escola Polit\'ecnica,  Universidade Federal do Rio de Janeiro,
C.P. 68529, 21941-909, Rio de Janeiro, RJ, Brazil,\\
$^2$ \'Ecole CentraleSupélec, Plateau du Moulon – 3, rue Joliot-Curie – 91192 GIF-SUR-YVETTE - France, \\
$3$ Paris Saclay, Espace Technologique, Bat. Discovery - RD 128 - 2e ét., 91190 Saint-Aubin, France,\\
$^4$ Instituto de F\'{\i}sica, Universidade Federal do Rio de Janeiro, CP 68528,
Rio de Janeiro, Brazil, \\
$^5$ Instituto de F\'{\i}sica, Universidade Federal Fluminense, Av. Litoranea
s/n, Gragoat\'{a}, Niter\'{o}i, R.J., 24210-340, Brazil, \\
$^6$ Instituto de Estudos Avan\c{c}ados, Universidade de S\~{a}o Paulo C. P.
72012, 05508-970 S\~{a}o Paulo-SP, Brazil, and Instituto de F\'{\i}sica,
Universidade de S\~{a}o Paulo, C. P. 66318, 05314-970 S\~{a}o Paulo,-SP,
Brazil, \\
$^7$ Departamento de F\'isica Matem\'atica, Instituto de F\'isica, Universidade de S\~ao Paulo, C.P. 66318, 05314-970, 
S\~ao Paulo, SP, Brazil, \\
$^8$ Departamento de F\'{i}sica, Instituto Tecnol\'{o}gico de Aeron\'{a}utica, CTA, S\~{a}o Jos\'{e} dos Campos, 
S\~ao Paulo, SP, Brazil
}

\date{\today}

\abstract{In this paper we revisit the one-dimensional tunnelling problem. We consider Kemble's approximation for the transmission 
coefficient. We show how this approximation can be extended to above-barrier energies by performing the analytical continuation of the 
radial coordinate to the complex plane. We investigate the validity of this approximation by comparing their predictions for the cross 
section and for the barrier distribution with the corresponding quantum mechanical results. We find that the extended Kemble's approximation 
reproduces the results of quantum mechanics with great accuracy.} 


\authorrunning{A.J. Toubiana \textit{et al.}}

\titlerunning{An improved WKB approximation for heavy ion fusion ...}

\maketitle


\section{Introduction}

Quantum tunneling (QT) has been with us for as long as the Schr\"odinger equation (SE). It describes a purely quantum 
phenomenon of a passage through a barrier. One of the first to popularize the essentials of QT was Gamow in his description 
of $\alpha$ decay of atomic nuclei. For this purpose, Gamow used an approximation of the solution of the SE based on 
the short wave length limit. This approximation came to be known as the WKB approximation and it corresponds to using 
the classical action as the phase of the wave function. In the development of the tunneling theory, several authors introduced 
concepts and recipes in order to get the tunneling probability as close as possible to the exact result obtained from the 
solution of the SE (for a recent review, see Ref.~\cite{HaT12}). \\

One such attempt was made by Kemble~\cite{Kem35} who introduced a form for the probability which works at energies below the top of the barrier, 
attains the value 1/2 at the top of the barrier as the exact result dictates, but fails at energies above the barrier. Hill and Wheeler~\cite{HiW53}, used a 
parabolic form of the barrier which allows an exact analytical solution of the Schr\"odingier equation, which coincides with the result obtained using the 
Kemble formula for the same barrier for energies below the top of the barrier. The HW form does work both below and above the barrier. However it 
has the major shortcoming that the parabolic approximation is only valid at energies in the vicinity of the top of the Coulomb barrier in nucleus-nucleus
collisions. \\

In the current paper we discuss the approximation of the Coulomb barrier by a parabola, which is frequently adopted in the description of fusion of nuclei. 
In this case, the tunneling probability is called the transmission coefficient. The purpose of this paper is to demonstrate that the general 
WKB-based Kemble formula for the tunneling probability can be extended to the classically allowed region encountered at energies above the top of the barrier.
With this extension we have a WKB formula for the tunneling probability valid for any well-behaved, complex nuclear potential at all energies.
The paper is organized as follows. In section II, we discuss fusion reactions in heavy ions collisions. In Section III we discuss the WKB approximation 
as used in nuclear physics, and introduce the concept of fusion barrier distributions, which has attracted considerable interest over the last two decades. 
In section IV, we develop the extension of the Kemble formula to energies above the barrier top. This development is made for a typical potential for
nucleus-nucleus collisions.  In section V, we use the approximations
discussed in the previous section in the calculations of fusion cross sections and barrier distributions in the case of the light
heavy ion system, $^6{\rm Li}+^{12}{\rm C}$. We chose a light system to stress the limitation of the parabolic approximation
discussed in the present work. For heavy systems, this approximation works better. They do break down but this occurs
at lower energies (in comparison to the Coulomb barrier). 
Finally, in Section VI several concluding remarks are presented.


\section{Fusion reactions in heavy ion collisions}

In a fusion reaction the projectile merges with the target, forming a compound nucleus (CN).
The densities of the collision partners overlap so strongly that they  lose their initial identities. 
Thus, the fraction of the incident current that reaches small projectile-target distances does 
not emerge in the elastic channel. In this way, any description of the collision based exclusively
on the elastic channel violates the continuity equation. In multi-channel descriptions of the collision, 
this is not a problem because the current removed from the elastic channel re-emerges (after 
a time-delay) when the CN decays. However, if one adopts a single-channel approach,
it is necessary to simulate the loss of flux in some way. This could be achieved
through the introduction of a negative imaginary part in the projectile-target interaction potential, 
$W^{\scr F}(r)$, or by adopting an ingoing wave boundary condition (IWBC). \\

In a heavy ion collision the real part of the potential has the general form,
\begin{equation}
V_l(r) = V_{\scr C}(r) + V_{\scr N}(r) + \frac{\hbar^2}{2\mu\, r^2}\ l(l+1),
\label{Veff}
\end{equation}
where  $\mu$ is the reduced mass of the projectile-target system, and $V_{\scr C}$ and  $V_{\scr N}$ are respectively the Coulomb and the nuclear potentials.
The third term in Eq.~(\ref{Veff}) is the centrifugal barrier, which appears in the radial equations.
For the sake of definiteness, we consider the $^6{\rm Li} - ^{12}$C collision throughout this paper.\\

As it is frequently done, we approximate the Coulomb potential as, 
\begin{eqnarray}
V_{\scr C}(r) &=&  \frac{ Z_{\scr P} Z_{\scr T}\,e^2}{r},
\qquad\qquad\qquad\ \ \  \ {\rm for\ } r\ge R_{\scr C}, \label{Vcou}
\\
                     &=& \frac{Z_{\scr P} Z_{\scr T}\,e^2}{2\,R_{\scr C}}\ \left( 3- \frac{r^2}{R_{\scr C}^2} \right), \label{Vcou-1}
                     \qquad {\rm for\ } r < R_{\scr C}.
                     \label{Vcou-2}
\end{eqnarray}
Above, $Z_{\scr P}$ and $Z_{\scr T}$ are respectively the atomic numbers of the projectile and the target, and
$e$ is the absolute value of the electronic charge. The {\it Coulomb radius}, $R_{\scr C}$,  is expressed in terms 
of the mass numbers of the projectile and the target ($A_{\scr P}$ and $A_{\scr T}$) as $R_{\scr C} = r_{0{\scr C}}\,
 \left( A_{\scr P}^{\scr 1/3}+ A_{\scr T}^{\scr 1/3}\right)$, with $r_{0{\scr C}}\simeq 1$ fm.\\

 For $V_{\scr N}(r)$, we adopt the Aky\"uz-Winther potential~\cite{BrW91}. This potential is evaluated through a double 
 folding integral of the nuclear densities with a M3Y~\cite{BBM77} nucleon-nucleon interaction~\cite{CaH13}. For practical
 purposes, it is approximated by the Woods-Saxon  (WS) function
 \begin{equation}
 V_{\scr N}(r) = \frac{V_0}  { 1+\exp\left[  \left( r - R_0 \right)/a_{\scr 0}  \right]  },
 \label{Vnuc-WS}
 \end{equation}
where $R_{0} = r_{0}\,  \left( A_{\scr P}^{\scr 1/3}+ A_{\scr T}^{\scr 1/3}\right)$. The WS parameters of the Aky\"uz-Winther
potential are expressed as functions of the mass numbers of the collision partners. For the $^6{\rm Li} + ^{12}$C system 
they have the values: $V_0 = - 31.35$ MeV, $r_0 = 1.16$ fm and $a_{\scr 0}=  0.56$ fm.\\

\begin{figure}
\centering
\resizebox{0.45\textwidth}{!}{ 
\includegraphics{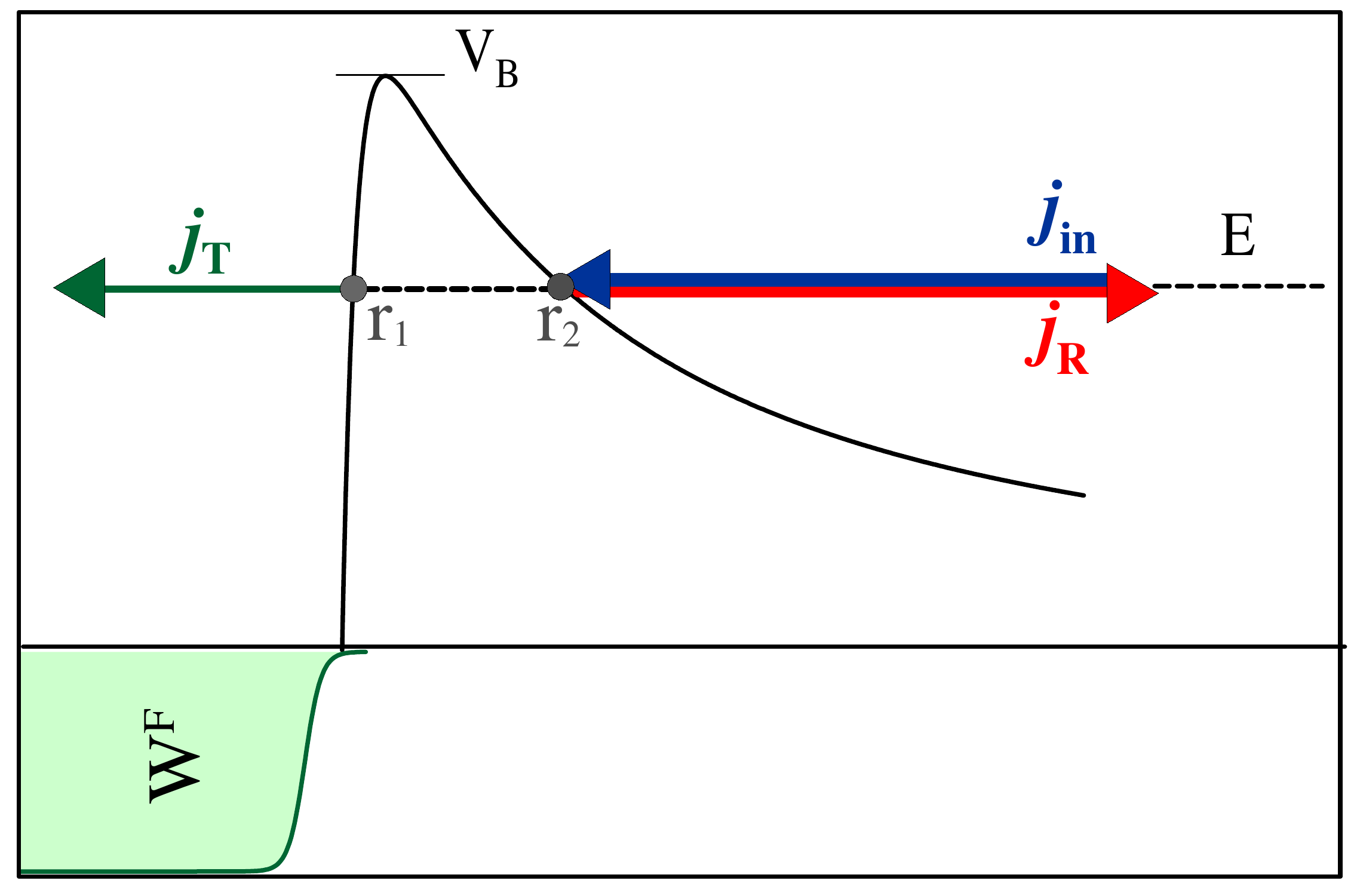}}
\caption{(Color online) Schematic representation of the real and the imaginary potentials in a typical heavy ion collision.
The classical turning points are indicated by $r_1$ and $r_2$.}
\label{potentials}
\end{figure}

In single-channel descriptions of fusion reactions based on Quantum Mechanics, one frequently uses a complex potential (the 
{\it optical potential}),  whose real part is given by Eq.~(\ref{Veff}). The sum of the attractive nuclear potential with the 
repulsive Coulomb and centrifugal potentials gives rise to a potential barrier, with height $V_{\scr B}$. The potential has 
also a negative imaginary  part, $W^{\scr F}$, very intense and with a short range, that accounts for the incident flux lost 
to the fusion channel. The situation is schematically represented in  Fig.~\ref{potentials}, for a collision energy ($E$)
below the Coulomb barrier. The figure shows the incident  current,  $j_{\rm in}$, the reflected current, 
$j_{\scr R}$,  and the current transmitted through the barrier, $j_{\scr T}$,  which reaches the strong absorption region. 
For practical purposes, the radial equation for each angular momentum is solved numerically. The numerical integration
starts at $r=0$, where the wave functions vanish and their derivatives are chosen arbitrarily. This choice only sets the 
normalization of the wave function, which has no influence on the cross sections. \\

Fusion absorption can also be simulated by an ingoing wave boundary condition (IWBC), with a real potential. In the 
IWBC~\cite{Raw64,Raw66,HRK99} one makes the assumption that the radial wave function at some distance 
$r=R_{\rm in}$ located in the inner region of the barrier (usually the minimum of $V_l(r)$) behaves as a wave propagating 
towards the origin. One then uses the WKB approximation to evaluate the radial wave function and its derivative at $r=R_{\rm in}$, 
and starts the numerical integration from this point. \\

The fusion cross section, can be evaluated by the partial-wave series
 \begin{equation}
 \sigma_{\scr F} = \frac{\pi}{k^2}\ \sum_{l=0}^\infty \left( 2l+1\right) \ P^{\scr F}_l P^{\scr CN}_l
 \label{lsumsigF}
 \end{equation}
where $P^{\scr F}_l$ is the absorption probability at the $l^{\rm th}$ partial-wave, given by
 \begin{equation}
P^{\scr F}_l = 1-\left| S_l \right|^2,
 \label{PFl}
 \end{equation}
and $P^{\scr CN}_l$ is the probability for the formation of the composite system, the compound nucleus, once the barrier is overcome. In most applications 
this probability is set equal to unity as the density of states of the CN is quite large. In light systems, such as $^{12}$C +$^{12}$C, it was demonstrated by 
Jiang {\it et al.}~\cite{JBE13} that at low energies the compound nucleus formation probability is significantly smaller than unity due to the low density of 
states of the CN, $^{24}$Mg at the excitation energies involved. However, this phenomenon lies beyond the scope of the present paper. Thus, we will set 
$P^{\scr CN}_l$ = 1 for all values of the orbital angular momentum.\\

Owing to the imaginary potential, or to an IWBC, the S-matrix is not unitary. Instead, it satisfies the condition:
$\left| S_l \right|^2 \le 1$. The deviation from unitarity, $1-\left| S_l \right|^2$, measures the absorption (fusion) probability.  
If absorption is simulated by the IWBC,  there is no outgoing wave for $0 < r <R_{\rm in}$. This means that the fraction of the incident current that 
reaches this region leads to fusion. Therefore, the fusion probability is given by the transmission coefficient through the potential barrier. 
Namely,
\begin{equation}
P^{\scr F}_l \simeq T_l = \frac{j_{\scr T}}{j_{\rm in}}.
\label{PlTl}
\end{equation}
\begin{figure}
\centering
\resizebox{0.45\textwidth}{!}{ 
\includegraphics{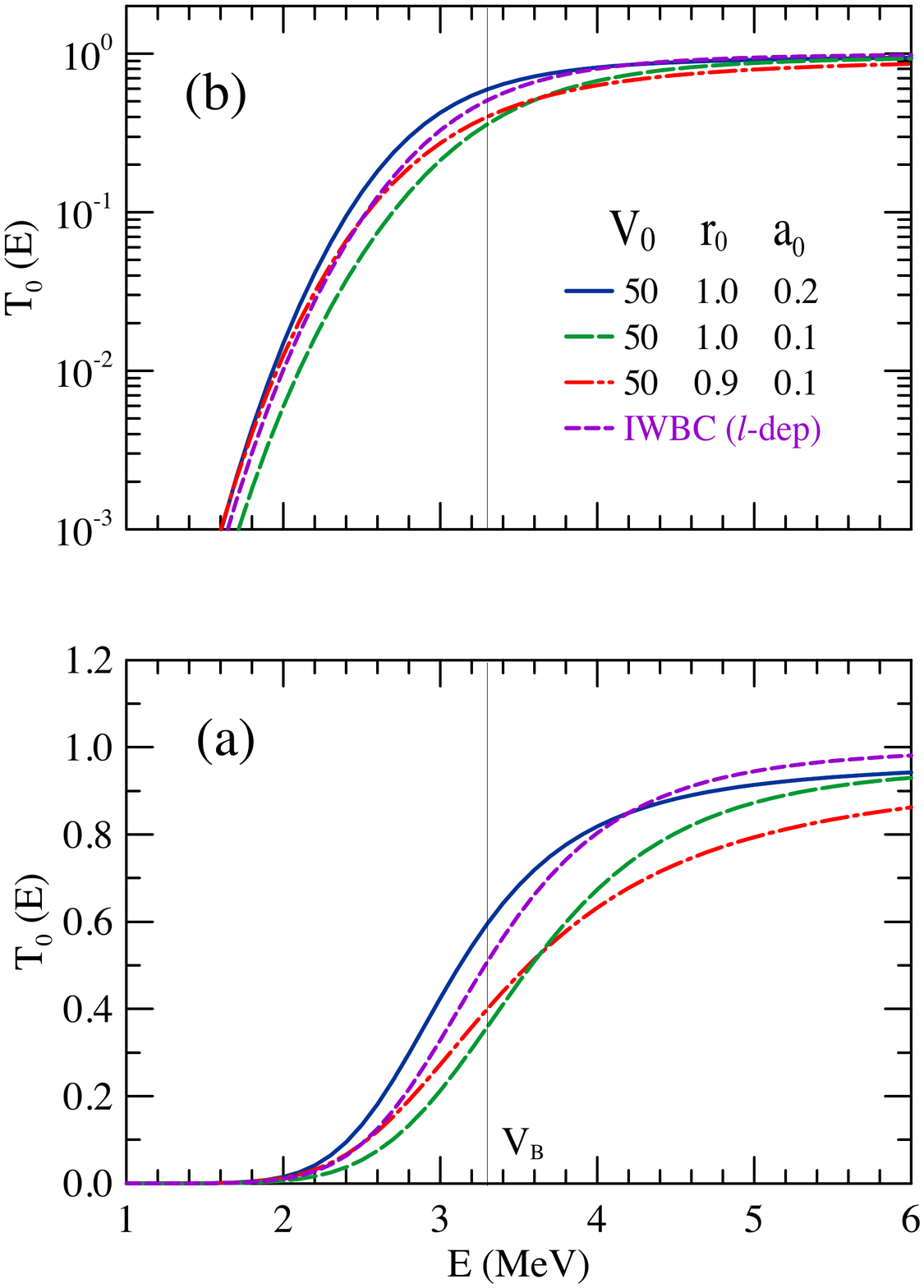}}
\caption{(Color online) The transmission coefficient obtained with the IWBC and fusion probabilities evaluated
with WS imaginary potentials with different $r_{\scr 0}$ and $a_{\scr 0}$ parameters (see the legend). The depth of the
imaginary potential is kept fixed at $V_0 = 50$ MeV. The results are shown in linear (panel (a)) and  logarithmic
(panel (b)) scales.}
\label{complexV-IWBC}
\end{figure}

\bigskip

In principle, the IWBC is equivalent to the effects of a deep imaginary potential acting exclusively in the inner region 
of the barrier. One frequently assumes that the details of the imaginary potential do not have significant influence on fusion
probabilities, as long as the potential is strong and has a short range. In fact, this assumption is questionable. It may be
valid for heavier systems but it is not accurate in collisions of light heavy ions, like $^6{\rm Li}\, -\, ^{12}{\rm C}$. This fact 
is illustrated in Fig.~\ref{complexV-IWBC}, where the s-wave fusion probabilities obtained with the IWBC and with 
frequently adopted values of the radius parameter and the diffusivity are shown.\\

Besides studying the energy dependence of the fusion cross section, it was found useful to go further and study the quantity
\begin{equation}
\mathcal{D}_{\scr F}(E)=\frac{d^{2}\left[E\sigma_{F}(E)\right]  }{dE^{2}},
\label{bardist-def}
\end{equation}
which is referred to as the {\it barrier distribution}~\cite{RSS91,HaR04}. Evaluating the second derivative numerically, experimental barrier distributions 
can be extracted directly from the data and comparing them with barrier distributions resulting from coupled channel calculations involving different 
channels, one can obtain invaluable information about the reaction mechanisms involved in the collision~\cite{DHR98,CGD06,CGD15}.

\subsection{The Wong formula and barrier distribution}

It has been a common practice in nuclear physics to use approximate forms of the potential barriers for the purpose of obtaining an analytical
solution of the scattering Schr\"odinger equation using the IWBC, and thus getting an analytical form for the transmission coefficient. This has 
been first done by Hill and Wheeler~\cite{HiW53}, who approximated
\[
V(r) = V_{\scr B} - \frac{1}{2}\,\mu\omega\ \left(r-R_{\scr B} \right)^2,
\]
In the above equations, $R_{\scr B}$ is the barrier radius, $V_{\scr B}$ is the barrier height and $\hbar\omega$ is the curvature parameter,
\begin{equation}
\hbar\omega = \sqrt{
\frac{ - \, \hbar^2\,V^{\prime\prime}\left( R_{\scr B}\right)} {\mu}}.
\label{HW}
\end{equation}
The transmission coefficient through this barrier, known as the Hill-Wheeler transmission coefficient, is given by the exact expression
\[
T^{\scr HW}(E) = \frac{1}{1+\exp\left[ 2\,\Phi^{\scr HW}(E) \right]} ,
\]
with
\[
\Phi^{\scr HW}(E) = \frac{2\pi}{\hbar\omega}\left(V_{\scr B} -E \right).
\]

This approximation could be used for each partial wave, leading to angular momentum dependent barrier radii, heights and curvatures. In this way,
one would have exact transmission coefficient for each $l$ and, using them in the partial-wave expansion, one could obtain the fusion cross section.\\

Back in 1973, Wong \cite{Won73} went one step further. He neglected the $l$-dependences of the barrier radii and curvatures, taking for all angular momenta
the values for $l=0$, denoted by $R_{\scr B}$ and $\hbar\omega$. In this way, the parabolic approximation for the effective potential is
\begin{equation}
V_l(r) = B_l - \frac{1}{2}\,\mu\omega\ \left(r-R_{\scr B} \right)^2 ,
\end{equation}
where $B_l$ is the barrier of the effective potential for the $l^{\rm th}$ partial-wave, given by
\begin{equation}
B_l =  V_{\scr B} +  \frac{\hbar^2}{2\mu R_{\scr B}^2}\ l(l+1).
\end{equation}
He then treated the angular momentum quantum number as a continuous variable, $l\rightarrow \lambda = l+1/2$, and approximated the sum of
partial-waves by an integral over $\lambda$. In this way, he got the closed expression,
\begin{equation}
\sigma^{\scr W}_{\scr F} (E) = \frac{\hbar\omega\, R_{\scr B}^2}{2E} \ln{\left[1 +\exp{\left(\frac{2\pi}{\hbar\omega} \left( E - V_{\scr B}\right)\right)}\right]}.
\label{wong}
\end{equation}
Eq. (\ref{wong}), known as the Wong formula, has been quite popular in low energy heavy-ion fusion and reactions. More recently, some 
improved versions of the Wong formula have been proposed~\cite{Won12,RoH15}. \\

\medskip

The fusion barrier distribution associated with the Wong formula can be evaluated easily. Using the cross section of Eq.~(\ref{wong}) in 
Eq.~(\ref{bardist-def}), one gets
\begin{equation}
\mathcal{D}_{\scr F}^{\scr W}(E)= 
\frac{\pi ^2R_{\scr B}^2/\hbar\omega}{1+\cosh \left[ 2\pi \left(E-V_{\scr B} \right) / \hbar\omega \right] }.
\label{WFBD-1}
\end{equation}
%


\section{The WKB approximation in Nuclear Physics}


The WKB approximation is a short wavelength limit of Quantum Mechanics. Since it is extensively 
discussed in text books on Quantum Mechanics and Scattering Theory~\cite{Mer98,Joa83,CaH13}, its
derivation will not be presented here. We consider only some applications in scattering theory.
We discuss the WKB approximation in the calculation of the transmission coefficients used to determine fusion cross 
sections (Eqs.~(\ref{lsumsigF}) and (\ref{PlTl})).\\ 

Let us consider the $^6{\rm Li}\ -\ ^{12}{\rm C}$ collision at a sub-barrier energy $E$, as represented in Fig.~\ref{potentials}.
Within the WKB approximation, the transmission coefficient through the barrier of the $l^{\rm th}$ partial-wave is given 
by
\begin{equation}
T_l^{\scr WKB} (E)= \exp\left[ -2\,\Phi_l^{\scr WKB}(E) \right].
\label{T0WKB}
\end{equation}
Above, $\Phi_l^{\scr WKB}$ is the integral 
\begin{equation}
\Phi_l^{\scr WKB}(E)  = \int_{r_1}^{r_2} \kappa_l(r)\ dr,
 \label{PhiWKB}
\end{equation}
where
\begin{equation}
\kappa_l(r) = \frac{\sqrt{2\mu\,\big[ V_l(r)-E\big] }}{\hbar}.
\label{kappa}
\end{equation}
In the above equations, $r_1$ and $r_2$ are the classical 
turning points, indicated in Fig.~\ref{potentials}. They are the solutions of the equation on $r$,
\begin{equation}
V_l(r) =E.
\label{turningpoints}
\end{equation}
Eq.~(\ref{T0WKB}) works very well for collision energies well below the Coulomb barrier, but it becomes very poor 
near the Coulomb barrier and above. At $E= B_l$, the two classical turning points coalesce, so that $\Phi_l^{\scr WKB}(E) =0$.
In this way, one gets  $T_l^{\scr WKB} (E) = 1$, whereas the quantum mechanical result is $T_l^{\scr QM} (E) = 1/2$.
Brink and Smilansky have shown that the situation can be improved if one takes into account multiple reflections under the
barrier~\cite{BrS83}.\\

In 1935, Kemble~\cite{Kem35} showed that the WKB approximation can be improved if one uses a better connection formula. 
He got the expression
\begin{equation}
T_l^{\scr K} (E)= \frac{1}{1+\exp\left[2\,\Phi_l^{\scr WKB}(E) \right]}.
\label{T0Kemble}
\end{equation}
At energies well below the Coulomb barrier, the two turning points are far apart and $\kappa(r)$ reaches appreciable values 
within the integration limits. In this way, $\Phi^{\scr WKB}(E)$ becomes very large, so that the unity can be neglected in the 
denominator of Eq.~(\ref{T0Kemble}). This equation then reduces to Eq.~(\ref{T0WKB}). Therefore,
the two approximations are equivalent in this energy region. On the other hand, Kemble's approximation remains valid as the
energy approaches the Coulomb barrier, so that at $E=B_l$ one gets $\Phi^{\scr WKB}(E) =1/2$, which is the correct result.\\

Unfortunately, Eq.~(\ref{T0Kemble}) gives wrong results at above-barrier energies. In this energy range, there are no classical turning points, 
since Eq.~(\ref{turningpoints}) has no real solutions. Thus, $\Phi^{\scr WKB}(E) = 0$. In this way, the transmission coefficient is frozen at 
the value 1/2. Therefore, it has the wrong high energy limit. 
However, Kemble~\cite{Kem35} pointed out that the validity of Eq.~(\ref{T0Kemble}) could be extended to 
above-barrier energies by an analytical continuation of the variable $r$ to the complex plane (see also Ref.~\cite{FrF65}),
although he did not elaborate on it. This point is considered in detail in the next section.


\section{Kemble transmission coefficient at above-barrier energies}


Eq.~(\ref{T0Kemble}) can be extended to above-barrier energies if one solves Eq.~(\ref{turningpoints}) in the complex r-plane and 
evaluates the integral of Eqs.~(\ref{PhiWKB}) and (\ref{kappa}) between the complex turning points. Complex turning points have
been previously used by different authors to evaluate WKB-phase shifts~\cite{BrT77,LTM78,LeT78,ViW89}. In this section, we apply this 
procedure to a typical potential in heavy ion scattering, and show that it actually works.\\

First, we introduce the dimensionless coordinate
 \begin{equation}
r\ \rightarrow\ x= \frac{r-R_{\scr B}}{a_{\scr 0}},
\label{new-coord}
 \end{equation}
where  $R_{\scr B}$ is the barrier radius and $a_{\scr 0}$ is the diffusivity parameter of the WS potential. Next, this 
coordinate is extended to the complex plane. That
is: $x\rightarrow z=x+i\,y$. The effective potential of Eq.~(\ref{Veff}) then becomes,
 \begin{multline}
V_l(z) = \frac{V_0}{1+\exp\left(z+d \right) } +
\frac{Z_{\scr P} Z_{\scr T}\,e^2}{a_{\scr 0} z+R_{\scr B}} \\
+ \frac{\hbar^2 \left(l+1/2 \right)^2}{2\mu\,\left( a_{\scr 0} z+R_{\scr B}\right)^2} .
\label{Vl-expl}
 \end{multline}
Above, 
 \begin{equation*}
d = \frac{R_{\scr B}- R_0}{a_{\scr 0}}
 \end{equation*}
is a dimensionless constant, associated with the difference between the barrier radius and the radius of the nuclear potential. Note that
in Eq.~(\ref{Vl-expl}) we replaced $l(l+1) \rightarrow (l+1/2)^2$. This is a common procedure in semiclassical theories. \\

Then, writing the now complex effective potential as,
\[ 
V_l(z)\equiv V_l(x+i\,y) = U_l(x,y) + i\,W_l(x,y),
\]
one imposes that its imaginary part vanishes. That is,
 \begin{equation}
 W_l(x,y) \equiv {\rm Im} \Big\{ V_l(z) \Big\} = 0.
 \label{W=0}
 \end{equation}
We should remind that the complexity of the potential is of purely mathematical origin and should not be confused with absorption of flux.
Using the explicit form of the effective potential (Eqs.~(\ref{Veff}), (\ref{Vcou}) and (\ref{Vnuc-WS})) in Eq.~(\ref{W=0}), one gets a complicated
equation, that describes lines on the complex z-plane. \\

\begin{figure}
\centering
\resizebox{0.45\textwidth}{!}{ 
\includegraphics{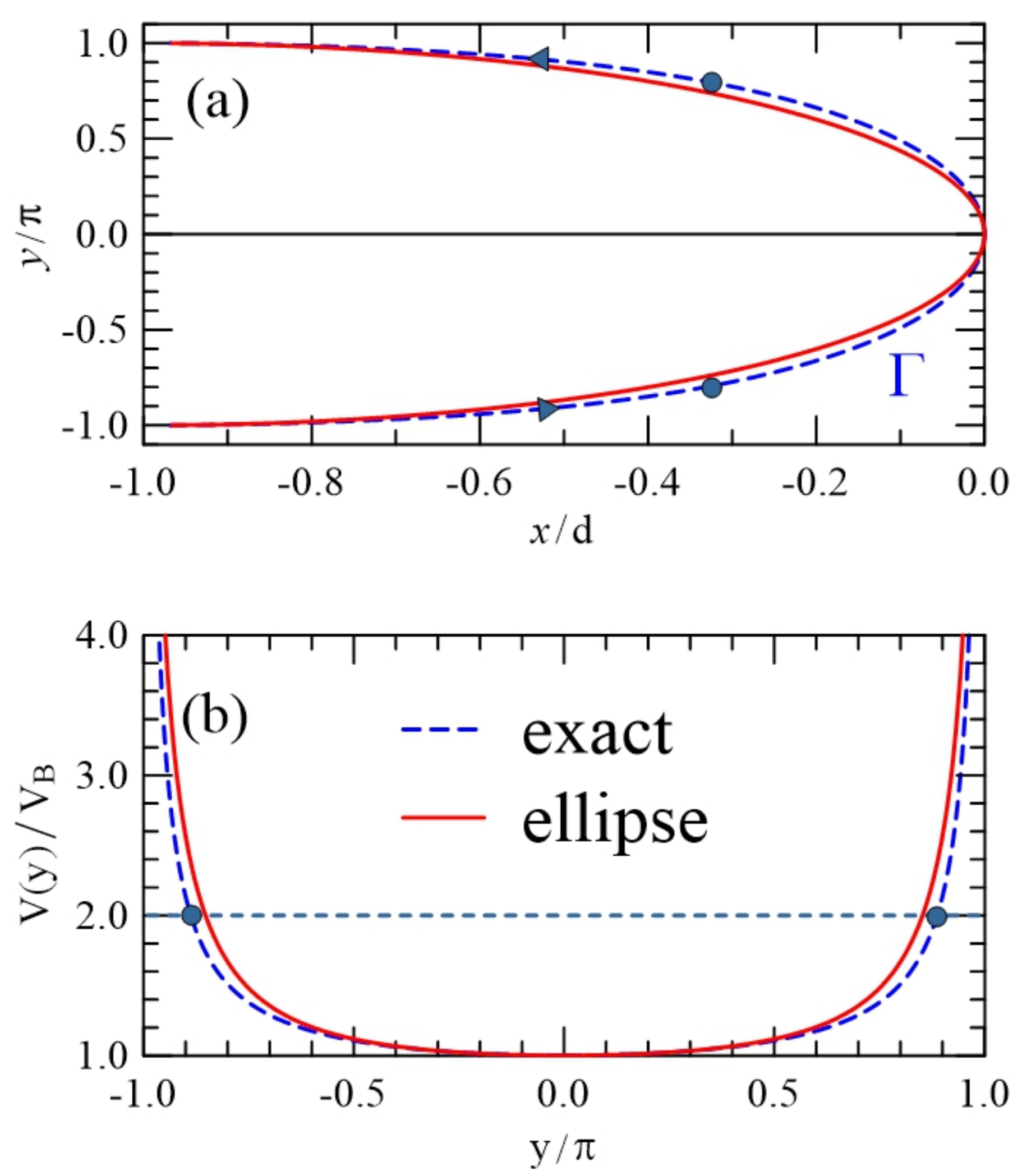}}
\caption{(Color online) The analytic continuation of the realistic potential of Eq.~(\ref{Veff}). Panel (a) shows the curves $\Gamma$ (blue dashed
line) and $\Gamma_{\rm e}$ (red solid line), and panel (b) shows the corresponding real potentials on these lines. For details, see the text.}
\label{Vl_z-plane}
\end{figure}

Solving Eq.~(\ref{W=0}) by a numerical procedure, one gets the curve $\Gamma$ represented by a blue dashed line on panel (a) of 
Fig.~\ref{Vl_z-plane}. We remark that there are other curves on the complex plane where the potential is real. 
First, there is a curve $\Gamma^\prime$, at
the left of $\Gamma$. That is, for $x< -d$. It looks like a distorted reflection of $\Gamma$ around a vertical axis at $x = -d$. 
In this way, joining this curve with $\Gamma$, one gets a closed contour. The potential is also real on other curves with $|y| > \pi$. The 
shapes of these curves are expected to be different from $\Gamma$, because the Coulomb and the centrifugal potentials are not periodic in $y$.
However, the physics involved in the collision is fully described by the analytical continuation on $\Gamma$. Therefore, we restrict our discussion
to this curve.

\medskip

We find that the curve $\Gamma$ is very close to the ellipse $\Gamma_{\rm e}$ given by the equation,
\begin{equation}
\left( \frac{x+d}{d} \right)^2 + \left( \frac{y}{\pi} \right)^2 = 1.
\label{ellipse}
\end{equation}
This ellipse is represented in Fig.~(\ref{Vl_z-plane}) by a red solid line. Clearly, the elliptical approximation for $\Gamma$ is indeed
quite good. 

\bigskip

Now we consider the potential over the contour $\Gamma$. Owing to Eq.~(\ref{W=0}), we can express $x$ as a function of $y$, $x_{\scr \Gamma}(y)$, 
so that the corresponding complex coordinate becomes,
\begin{equation}
z_{\scr \Gamma}(y) = x_{\scr \Gamma}(y) + i\,y. 
\label{z_Gamma}
\end{equation}
In this way, the potential depends exclusively on the coordinate $y$. To simplify the notation of the potential over $\Gamma$, where it is real, we write
\begin{equation}
U_{\scr \Gamma}(y) = V_l \left( z_{\scr \Gamma}(y)\right).
\label{U-z_Gamma}
\end{equation}
Since Eq.~(\ref{W=0}) is rather complicated, the function $x_{\scr \Gamma}(y)$ cannot be obtained analytically. Therefore, the potential
$U_{\scr \Gamma}(y)$ must be determined numerically. 

\medskip

The situation is much simpler within the elliptical approximation for $\Gamma$. Using Eq.~(\ref{ellipse}), we get 
\begin{equation*}
x_{\scr \Gamma_{\rm e}} (y) = -d\, \left[ 
1 - \sqrt{1-\frac{y^2}{\pi^2}}   \right] .
\end{equation*}
Inserting the above equation into Eq.~(\ref{z_Gamma}), and using the result in Eq.~(\ref{Vl-expl}), we get an analytical expression for the
potential $V_l \left( z_{\scr \Gamma_{\rm e}}(y)\right)$. However, this potential is not exactly real. Owing to the approximation of 
$\Gamma$ by $\Gamma_{\rm e}$, $V_l \left( z_{\scr \Gamma_{\rm e}}(y)\right)$ has a small imaginary
part, which should be discarded. We then write,
\begin{equation}
U_{\scr \Gamma_{\rm e}}(y) = {\rm Re} \Big\{V_l \left( z_{\scr \Gamma_{\rm e}}(y)\right) \Big\}.
\label{z_Gamma-1}
\end{equation}
This potential, evaluated for $l=0$, corresponds to the red solid line in Fig.~(\ref{Vl_z-plane}) (b). Clearly it is very close to the exact potential.

\medskip

Inspecting $U_{\scr \Gamma}(y)$, one immediately notices that it goes to infinity at the two extremes of the plot, which correspond to 
$y=\pm \pi$. These divergences can be traced back to the poles of the WS potential, located at $r=R_0 + i n\pi$, where $n$ is
any positive or negative integer. In terms of $z$, the poles are located at $z=-\, d+ i\, n \pi$. 
The infinite value of the potential at $y = \pm \pi$ has two important consequences. The first is that it 
confines the complex plane to the region $ -\pi < y < \pi$, which corresponds to $-d < x < 0$. In this way, the curve $\Gamma^\prime$ and curves 
with $| y | >\pi$ are not accessible to the system. Thus, such curves can be completely ignored. The second consequence is that the minimal
value of $x$ on $\Gamma$ is $x = -d$, which corresponds to $r_{\rm min} = R_0$ and this value is larger than $R_{\scr C}$. Therefore, the
Coulomb potential for points on $\Gamma$ are always given by Eq.~(\ref{Vcou}). This justifies our choice of the Coulomb potential in 
Eq.~(\ref{Vl-expl}).\\

\medskip

From Fig.~\ref{Vl_z-plane} we conclude that ellipse is very close to  $\Gamma$, and the potentials over the two curves are nearly the same. 
Thus, we henceforth replace the contour $\Gamma$ by the approximate curve $\Gamma_{\rm e}$. This simplifies our calculations considerably.\\

Now we evaluate Kemble's transmission factor. For this purpose, one has to calculate the WKB integral between the complex turning points, $z_{\scr \pm}$,
\begin{equation}
\Phi^{\scr WKB}(E) =a_{\scr 0}\ \frac{\sqrt{2\mu}}{\hbar}\   \int_{z_{\scr -}}^{z_{\scr +}} 
dz\ \sqrt{V_l \left( z_{\scr \Gamma_{\rm e}} \right) - E }\, .
\label{int-z}
\end{equation}
We remark that the factor $a_{\scr 0}$ results from the change of variable of Eq.~(\ref{new-coord}). The integral of Eq.~(\ref{int-z}) is independent of the
integration path on the complex-plane. It depends only on the integration limits, $z_{\scr -}$ and $z_{\scr +}$. However, for practical purposes, 
it is convenient to evaluate the integral along the contour $\Gamma_{\scr e}$. 
On this contour, $V_l(x,y)$ reduces to the real potential 
of Eq.~(\ref{z_Gamma-1}), $U_{\scr \Gamma_{\rm e}}(y)$,  and the differential $dz$ can be written as
\begin{equation*}
dz = \left[\frac{dx_{\scr \Gamma_{\rm e}}(y)}{dy} + i\right]\ dy,
\end{equation*}
where $dx_{\scr \Gamma_{\rm e}}(y)/dy$ is a real function of $y$. 
For energies above the barrier, $U_{\scr \Gamma_{\rm e}}(y) - E$ is negative so that,
\begin{eqnarray}
dz\ \sqrt{U_{\scr \Gamma_{\rm e}}(y) -E} &=& \left( \frac{dx_{\scr \Gamma_{\rm e}}(y)}{dy} + i\right) \times i\ \sqrt{E-U_{\scr \Gamma_{\rm e}}(y)} \ dy \nonumber\\
                                                    &=&  G_{\rm R}(y) \ dy + i\, G_{\rm I}(y)\ dy,
                                                    \label{GR-GI-1}
\end{eqnarray}
where $G_{\rm R}(y)$ and  $G_{\rm I}(y)$ are the real functions,
\begin{eqnarray}
G_{\rm R}(y) &=&  - \sqrt{E-U_{\scr \Gamma_{\rm e}}(y)} \label{GR}, \\
G_{\rm I}(y) &=&  \frac{d x_{\scr \Gamma_{\rm e}}(y)}{dy}\ \, \sqrt{E-U_{\scr \Gamma_{\rm e}}(y)} .
\label{GI}
\end{eqnarray}

\medskip

The values of $y$ corresponding to the integration limits of Eq.~(\ref{int-z}) are given by the equation,
\begin{equation*}
U_{\scr \Gamma_{\rm e}}(y) = E.
\end{equation*}
They are schematically represented in Fig.~(\ref{Vl_z-plane}), for the collision energy $E=2\,V_{\scr B}$. Since the potential $U_{\scr \Gamma_{\rm e}}(y)$
is symmetric with respect to $y=0$, the turning points have the property,
\begin{equation}
y_{\scr -} =- y_{\scr +} .
\label{y_pm}
\end{equation}
Thus, Eq.~(\ref{y_pm}) indicates that the integration limits are symmetric.
This property can be formally proved. Since $F\left( z \right) \equiv E - V_l\left( z \right)$ is an analytic function of $z$ in the vicinity of $\Gamma$
(or $\Gamma_{\rm e}$), we can write $F^*\left( z \right) = F\left( z^* \right)$. Thus, if $z_{\scr +}$ is a complex turning point, that is, it satisfies the 
equation $F\left( z_{\scr +} \right) = 0$, $z_{\scr -} = z^*_{\scr +}$ will also be a turning point. Then, writing $z_{\scr +} = x_{\scr +} +i\,y_{\scr +}$,
the other turning point, will be $z_{\scr -} = x_{\scr -} +i\,y_{\scr -}$, with $x_{\scr -}=x_{\scr +}$ and  $y_{\scr -}=-\,y_{\scr +}$.

\medskip

Expressing the integral of Eq.~(\ref{int-z}) in terms of the variable $y$, one gets
\begin{multline}
\Phi^{\scr WKB}(E) =a_{\scr 0}\, \frac{\sqrt{2\mu}}{\hbar}\  {\rm Re}\Bigg\{ \int_{y_{\scr -}}^{y_{\scr +}} G_{\rm R}(y)\ dy\, \\
+\, i \int_{y_{\scr -}}^{y_{\scr +}} G_{\rm I}(y)\ dy \Bigg\}.
\label{GR-GI}
\end{multline}
Fig.~\ref{Vl_z-plane}(a)  indicates that the potential $U_{\scr \Gamma_{\rm e}}(y)$ is an even function of $y$, and so is the function $G_{\rm R}(y)$ of  
Eq.~(\ref{GR}).  On the other hand, inspecting Fig.~\ref{Vl_z-plane}(b) one concludes that $dx_{\Gamma_{\rm e}}(y)/dy$ is an odd function of $y$. 
For this reason, the function $G_{\rm I}(y)$ of  Eq.~(\ref{GI}) is odd. Since the integration limits in Eq.~(\ref{GR-GI}) are symmetric, the second integral 
in this equation vanishes. Then, the WKB integral reduces to,
\begin{equation*}
\Phi^{\scr WKB}(E) = - a_{\scr 0}\, \frac{\sqrt{2\mu\,V_{\scr B} }}{\hbar}\ \int_{y_{\scr -}}^{y_{\scr +}}  
dy\ \sqrt{\varepsilon - \frac{U_{\scr \Gamma_{\rm e}}(y)}{V_{\scr B}}}.
\label{Int-T0-Vexact}
\end{equation*}
Above, we have introduced the notation,
\begin{equation*}
\varepsilon = \frac{E}{V_{\scr B}}.
\end{equation*}
Inserting Eq.~(\ref{Int-T0-Vexact}) into (Eq.~(\ref{T0Kemble})), one gets Kemble's transmission coefficient at above-barrier energies.


\section{Study of fusion in the $^6{\rm Li} - ^{12}{\rm C}$ collision}

%

\begin{figure}
\centering
\resizebox{0.40\textwidth}{!}{ 
\includegraphics{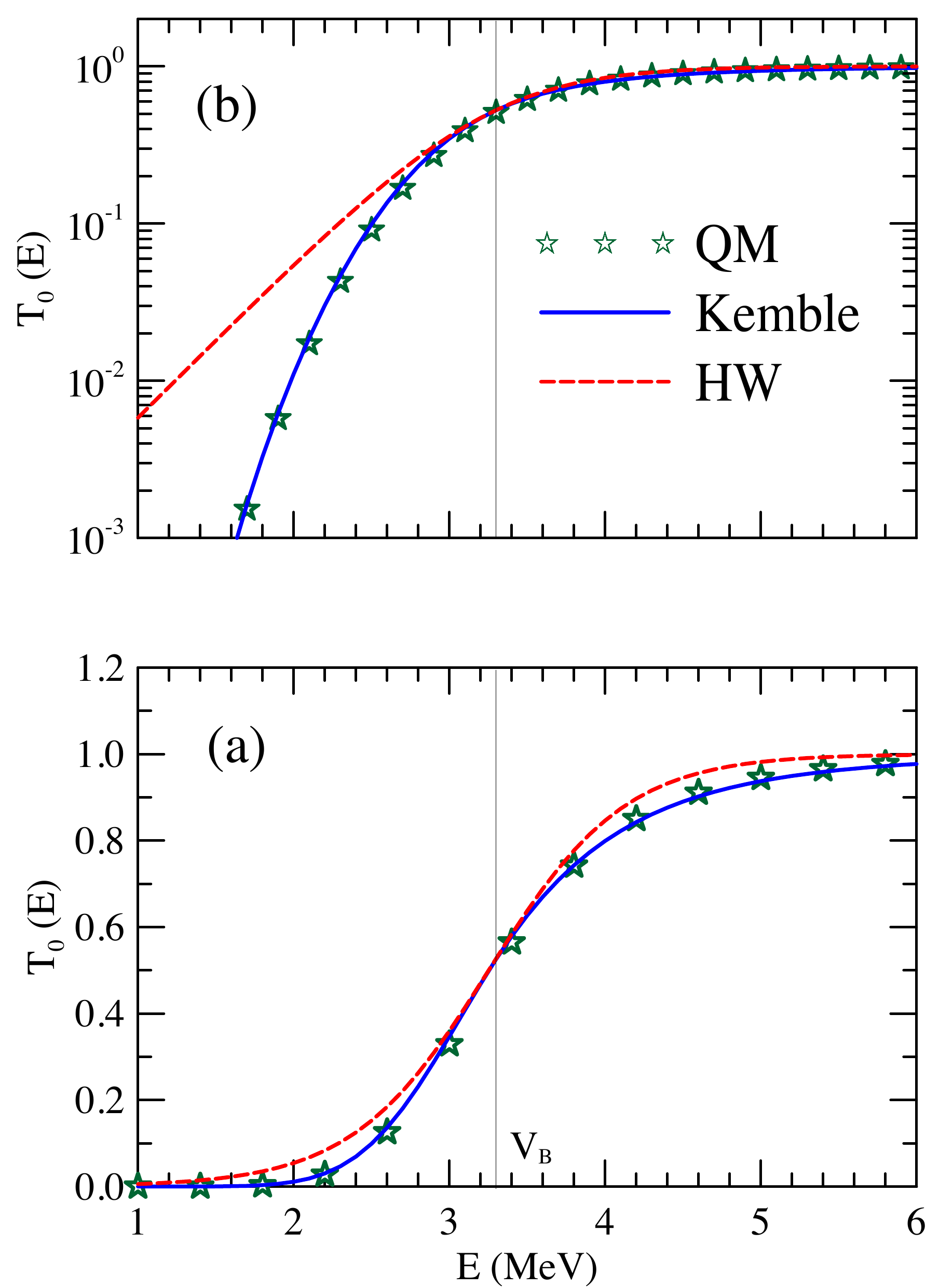}}
\caption{(Color online) Comparison the S-wave transmission coefficients obtained with exact quantum mechanics (blue solid lines), 
with the Kemble's approximation (stars), and the Hill-Wheeler transmission coefficient of Eq.~(\ref{HW}) (red dashed-lines), corresponding 
to a parabolic approximation to the Coulomb barrier. The system is $^{6}$Li +$^{12}$C. The results are shown in linear (panel (a)) and logarithmic (panel (b)) scales.}
\label{T0-fig}
\end{figure}

In this section we assess the validity of the Kemble approximation extended to the complex $r$-plane, discussed in the previous
section. In Fig. \ref{T0-fig} we compare the S-wave transmission coefficients obtained with this approximation (blue solid lines) with 
the corresponding quantum mechanical results (stars). The figure shows also the Hill-Wheeler transmission coefficients, 
corresponding to the parabolic fit to the Coulomb barrier (red dashed lines). Clearly, the Kemble approximation reproduces 
accurately the exact results, above and below the Coulomb barrier. On the other hand, the Hill-Wheeler transmission coefficients 
are close to the exact ones at energies above the barrier, but it overestimates them by orders of magnitude at sub-barrier energies.
This problem can be traced back to the parabolic approximation to the Coulomb barrier at large radial distances. Although the parabola
is close to the Coulomb barrier at $r\sim R_{\scr B}$, it falls off much faster as $r$ increases. Then, at low collision energies, the classical
turning point for the parabolic barrier is much larger than the one for the exact potential. This makes the Hill-Wheeler transmission 
coefficient unrealistically large.\\

\begin{figure}
\centering
\resizebox{0.40\textwidth}{!}{ 
\includegraphics{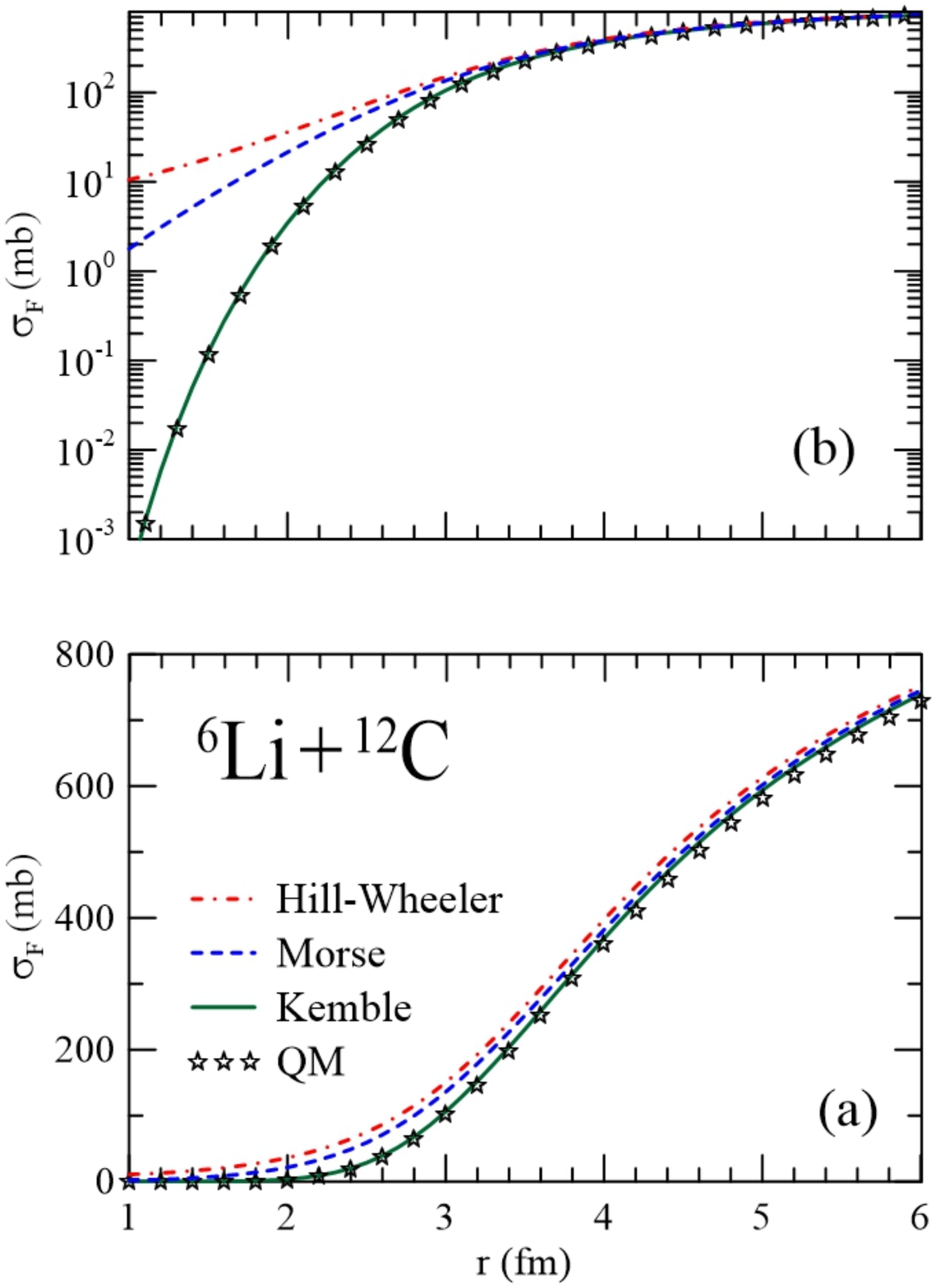}}
\caption{(Color online) Comparison of the exact quantum mechanical fusion cross section for the $^6{\rm Li}+^{12}{\rm C}$ systems with
results of approximations discussed in the previous sections. The results are shown in linear (panel (a)) and logarithmic (panel (b)) scales.}
\label{sigfus-fig}
\end{figure}

Now we discuss the use of the same approximations to predict observable quantities. First we look at fusion cross sections.
We compare the exact cross section, obtained with full quantum mechanics (stars), with the cross section using Kemble's transmission
coefficients through the exact barriers (blue solid lines). In all Kemble's calculations, we used the elliptical approximation for $\Gamma$. 
We consider also the Wong cross section, which is evaluated with exact transmission coefficients but through the parabolae
fitting the exact $l$-dependent barriers. The results are shown in Fig.~\ref{sigfus-fig}, in linear (panel (a)) and 
logarithmic (panel (b)) scales. 

\smallskip

One observes that the three cross sections of Fig.~\ref{sigfus-fig} show the same trends of the  S-wave transmission coefficients of
the previous figure. First, one notices that Kemble's approximation reproduces accurately the quantum mechanical cross section in the 
whole energy range. The cross section obtained with Wong's approximations is reasonably close to its quantum mechanical counterpart 
in the vicinity of the Coulomb barrier, and  it is slightly larger at higher energies ($E   >  4.5$ MeV). The reason is that, as the energy 
increases, Wong's cross section gets relevant contributions from higher partial-waves, where the radius and the curvature parameters 
become significantly different from those for $l=0$, used for all partial-waves. On the other hand, the situation is very bad at sub-barrier 
energies. The agreement with the quantum mechanical cross section is poor, becoming progressively worse as the energy decreases. 
At the lowest energies in the plot, $E\sim 1$ MeV, it overestimates the quantum mechanical cross section by more than three orders of 
magnitude. The problem with Wong's cross section in this energy region results from the parabolic approximation to the $l$-dependent barriers. As
we explained in the discussion of the previous figure, the transmission coefficients through parabolae for $E\ll B_l$ are much 
larger than those through the corresponding exact barriers.\\

\begin{figure}
\centering
\resizebox{0.45\textwidth}{!}{ 
\includegraphics{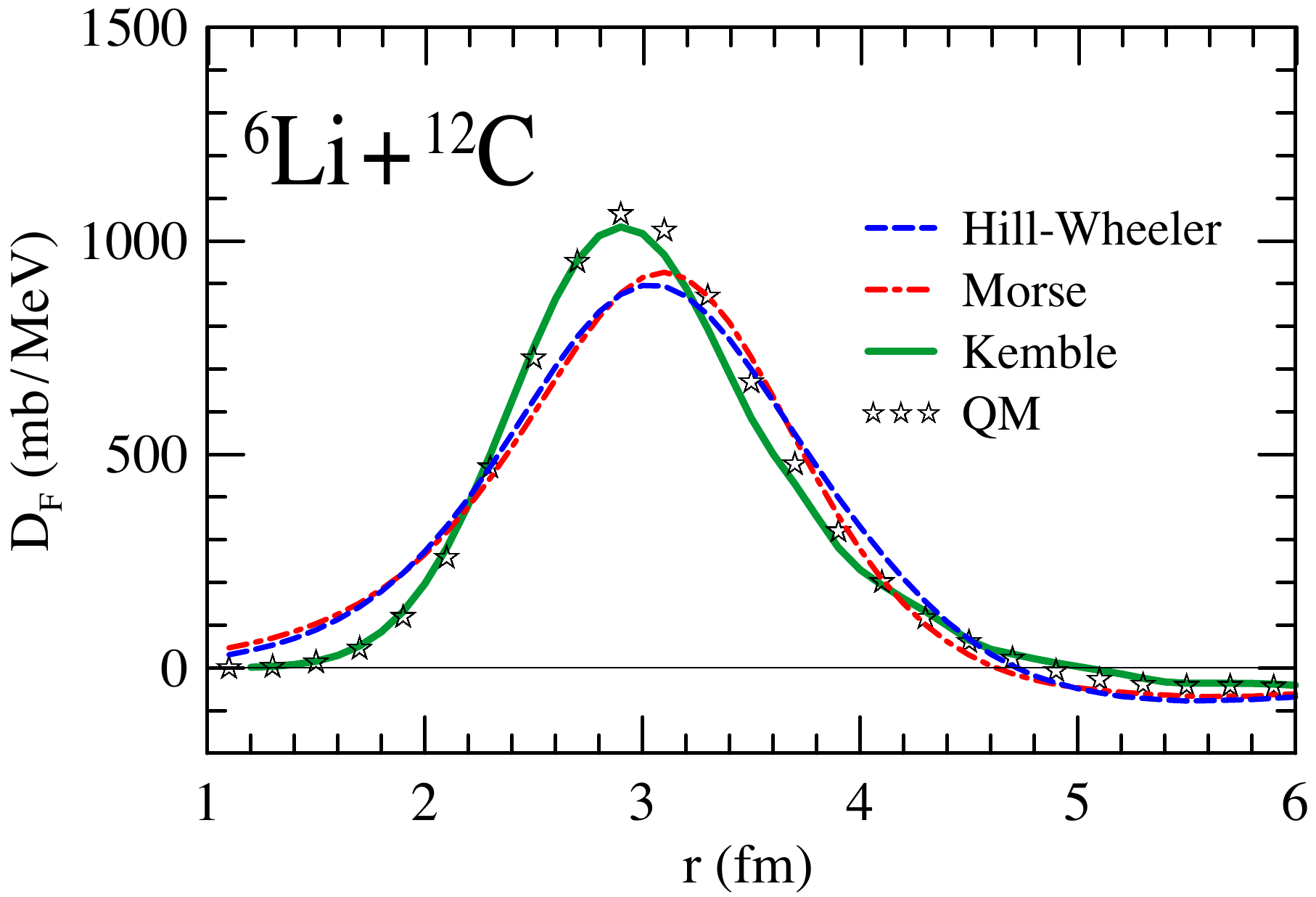}}
\caption{(Color online) Barrier discributions for the fusion cross sections of the previous figure.}
\label{bardist-fig}
\end{figure}

Now we discuss the predictions of the same calculations for the fusion barrier distributions. The results are presented in Fig.~\ref{bardist-fig}.
The notation is the same as in the previous figure. Since barrier distributions are always calculated near the barrier, Wong's barrier distribution is
not too bad. It is in qualitative agreement with the quantum mechanical fusion barrier distribution. On the other hand, the results of Kemble's WKB
are extremely good. They reproduce the exact barrier distribution throughout the energy region of the figure. An interesting aspect of the figure
is that the maximum of the quantum mechanical barrier, and also of Kemble's barrier distribution, occurs at $E_{\rm max} = 2.95$ MeV. Although is 
not far from the barrier of the bare potential ($V_{\scr B} = 3.25$ MeV), it is about 0.3 MeV lower.


\section{Conclusions}


In this paper we considered the more general, and potentially more useful, Kemble approximation for the 
tunnelling probability, which is an improved 
version of the WKB approximation. We considered in detail the extension of Kemble's approximation to energies above 
the barrier, which can be achieved by judicious contour integration in the complex $r$-plane. By a judicious choice of a contour in the complex r-plane we have extended the 
validity of the Kemble formula to energies above the barrier. We have accomplished this for barriers of a typical optical potential in heavy ion collisions.\\

We have discussed also cross sections and barrier distributions, as given by the second derivative of $E\sigma_{F}$ with 
respect to energy, and shown that Kemble's approximation gives an exceedingly good description of the quantum
mechanical results at near-barrier energies. Thus we have, in this paper, finally obtained a WKB tunneling probability which is valid for an arbitrary potential and at all energies, without relying on the parabolic or other approximation.

\bigskip

\textbf{
We are grateful to Kouichi Hagino for providing us with a single-channel code using the ingoing wave boundary condition, and to Raul Donangelo for 
critically reading the manuscript. Partial support from the Brazilian funding agencies, CNPq, FAPESP, and FAPERJ is also acknowledged. 
M. S. H. acknowledges support from the CAPES/ITA Senior Visiting Professor Fellowship Program A.J.T. acknowledges financial support from the 
French committee of the Brafitec program.}

\bibliographystyle{epj}

\begin{thebibliography}{26}

\bibitem{HaT12}
K.~Hagino, N.~Takigawa, Prog. Theor. Phys. \textbf{128}, 1061 (2012)

\bibitem{Kem35}
E.C. Kemble, Phys. Rev. \textbf{48}, 549 (1935)

\bibitem{HiW53}
D.L. Hill, J.A. Wheeler, Phys. Rev. \textbf{89}, 1102 (1953)

\bibitem{BrW91}
R.A. Broglia, A.~Winther, \emph{Heavy Ion Reactions} (Westview Press, 2004)

\bibitem{BBM77}
G.F. Bertsch, J.~Borysowicz, H.~McManus, W.G. Love, Nucl. Phys. \textbf{A284},
  399 (1977)

\bibitem{CaH13}
L.F. Canto, M.S. Hussein, \emph{Scattering Theory of Molecules, Atoms and
  Nuclei} (World Scientific Publishing Co. Pte. Ltd., 2013)

\bibitem{Raw64}
G.H. Rawitscher, Phys. Rev. \textbf{135}, B605 (1964)

\bibitem{Raw66}
G.H. Rawitscher, Nucl. Phys. \textbf{85}, 337 (1966)

\bibitem{HRK99}
K.~Hagino, N.~Rowley, A.T. Kruppa, Comp. Phys. Commun. \textbf{123}, 143 (1999)

\bibitem{JBE13}
C.L. Jiang, B.B. Back, H.~Esbensen, R.V.F. Janssens, K.E. Rehm, R.J. Charity,
  Phys. Rev. Lett. \textbf{110}, 072701 (2013)

\bibitem{RSS91}
N.~Rowley, G.R. Satchler, P.H. Stelson, Phys. Lett. B \textbf{254}, 25 (1991)

\bibitem{HaR04}
K.~Hagino, N.~Rowley, Phys. Rev. \textbf{C69}, 054610 (2004)

\bibitem{DHR98}
M.~Dasgupta, D.~Hinde, N.~Rowley, A.~Stefanini, Ann. Rev. of Nucl. Part. Sci.
  \textbf{48}, 401 (1998)

\bibitem{CGD06}
L.F. Canto, , P.R.S. Gomes, R.~Donangelo, M.S. Hussein, Phys. Rep.
  \textbf{424}, 1 (2006)

\bibitem{CGD15}
L.F. Canto, P.R.S. Gomes, R.~Donangelo, J.~Lubian, M.S. Hussein, Phys. Rep.
  \textbf{596}, 1 (2015)

\bibitem{Won73}
C.Y. Wong, Phys. Rev. Lett. \textbf{31}, 766 (1973)

\bibitem{Won12}
C.Y. Wong, Phys. Rev. C \textbf{86}, 064603 (2012)

\bibitem{RoH15}
N.~Rowley, K.~Hagino, Phys. Rev. C \textbf{91}, 044617 (2015)

\bibitem{Mer98}
E.~Merzbacher, \emph{Quantum Mechanics}, 3rd~edn. (John Wiley \& Sons, Inc.,
  1998)

\bibitem{Joa83}
C.J. Joachain, \emph{Quantum Collision Theory} (North Holland, 1983)

\bibitem{BrS83}
D.M. Brink, U.~Smilansky, Nucl. Phys. A \textbf{405}, 301 (1983)

\bibitem{FrF65}
N.~Fr\"oman, P.~Fr\"oman, \emph{JWKB Approximation: Contributions to the
  Theory}, 1st~edn. (North-Holland, Amsterdam, 1965)

\bibitem{BrT77}
D.M. Brink, N.~Takigawa, Phys. Rev. \textbf{C47}, R2470 (1977)

\bibitem{LTM78}
S.Y. Lee, N.~Takigawa, C.~Marty, Nucl. Phys. A \textbf{308}, 161 (1978)

\bibitem{LeT78}
S.Y. Lee, N.~Takigawa, Nucl. Phys. \textbf{A308}, 189 (1978)

\bibitem{ViW89}
E.~Vigezzi, A.~Winther, Ann. Phys. \textbf{192}, 432 (1989)

\end{thebibliography}

%
%
%
%

\end{document}